\documentclass[11pt,a4paper]{article} 

\catcode`\@=11
\def\lsim{\mathrel{\mathpalette\@versim<}}
\def\gsim{\mathrel{\mathpalette\@versim>}}
\def\@versim#1#2{\vcenter{\offinterlineskip
\ialign{$\m@th#1\hfil##\hfil$\crcr#2\crcr\sim\crcr } }}
\catcode`\@=12

\catcode`@=12
\usepackage{jheppub}
\usepackage[export]{adjustbox}
\usepackage{float}
\usepackage{caption}
\usepackage{subcaption}
\usepackage{epsfig,multicol,bbm}
\usepackage{graphicx}
\usepackage{amsmath}
\usepackage{slashed}
\usepackage{tikz}
\usepackage{bm}
\usepackage{amssymb}

\begin{document}
\thispagestyle{empty}
\vspace{0.6in}
\begin{center}
{\LARGE \bf A Radiative Model of Quark Masses with Binary Tetrahedral Symmetry\\}
\vspace{1.2in}
{\bf Alexander Natale\footnote{alexnatale@kias.re.kr}\\}
\vspace{0.2in}

{\sl Korea Institute for Advanced Study (KIAS),\\
Seoul 02455, Republic of Korea\\}
\end{center}
\vspace{0.75in}
\begin{abstract}

A radiative model of quark and lepton masses utilizing the binary tetrahedral ($T^{\prime}$) flavor symmetry, or horizontal symmetry, is proposed which produces the first two generation of quark masses through their interactions with vector-like quarks that carry charges under an additional $U(1)$.  By softly-breaking the $T^{\prime}$ to a residual $Z_4$ through the vector-like quark masses, a CKM mixing angle close to the Cabibbo angle is produced.  In order to generate the cobimaximal neutrino oscillation pattern ($\theta_{13}\neq0,\theta_{23}=\pi/4,\delta_{CP}=\pm \pi/2$) and protect the horizontal symmetry from arbitrary corrections in the lepton sector, there are automatically two stabilizing symmetries in the dark sector.  Several benchmark cases where the correct relic density is achieved in a multi-component DM scenario, as well as the potential collider signatures of the vector-like quarks are discussed.  

\end{abstract}

\newpage
\baselineskip 24pt

\section{Introduction}
Neutrino mass and oscillation are thoroughly established experimentally~\cite{sno,sno2,kamland,kamland2,t2k,k2k,kam2,superk,doublechooz}, as is the strong evidence for cosmological dark matter (DM)~\cite{DMrev,wmap,Ade:2015xua}; both are widely considered the best evidence for physics beyond the Standard Model (BSM), particularly given the robustness of the Standard Model (SM) at explaining the 2016 Run at the Large Hadron Collider (LHC).  Neutrino mass, and subsequent oscillation, pose an interesting set of questions: why is there such a large gap in the scale of neutrino mass relative to the rest of the fermion masses in the SM, and why is the neutrino oscillation the way that it is --- comparitivaley large angles relative to the Cabibbo--Kobayashi--Maskawa (CKM) matrix?  A general framework, proposed in 2006, tries to solve these issues by generating neutrino mass radiatively through the interactions of neutrinos with DM at the one-loop level~\cite{scotoma1}. While this proposal was not the first model of radiative neutrino mass~\cite{Zee:1980ai}, or the first model that completed the loops with DM~\cite{Krauss:2002px}, these so-called scotogenic, or Ma, models provide a comparitively simple way to connect neutrino mass and DM with a single Higgs at the one-loop level.  These Ma models have also been extended to explain lepton and quark mass~\cite{Ma:2014yka}, which yields interesting signatures at colliders~\cite{acnscotowdm,Ma:2014wea,Fraser:2015zed}.  Recently, there has been interest in extending such radiative models through the addition of vector-like fermions, and in particular vector-like quarks~\cite{Nomura:2016emz,Nomura:2016ask}. Additionally, a program utilizing various non-Abelian discrete flavor symmetries, or horizontal symmetries, has been pursued to explain the particular pattern of neutrino oscillation (see Refs.~\cite{King:2013eh,ishimorigroup} for reviews).  The measurement of $\theta_{13} \neq 0$~\cite{renotheta13nonzero,dayatheta13nonzero}, and the observation of a 125 GeV Higgs-like boson~\cite{higgsAtlas,higgsCMS}, disfavors many minimal models of horizontal symmetries that seek to explain the structure of the Pontecorvo--Maki--Nakagawa--Sakata (PMNS) matrix.  However, a recent proposal for a modified scotogenic model of neutrino mass with a tetrahedral ($A_4$) horizontal symmetry~\cite{MaCobimaximal} is able to support the so-called co-bimaximal mixing pattern as a genuine prediction where $\theta_{13} \neq 0$, $\delta_{CP} = \pm \frac{\pi}{2}$, and $\theta_{23}$ is maximal.  While $\theta_{23}$ being maximal is disfavored by NO$\nu$A at the 2.5 $\sigma$ level~\cite{Adamson:2016xxw}, the maximal value of $\theta_{23}$ is still consistent with most of the neutrino oscillation data~\cite{Timmons:2015laq,deAndre:2016cdt,Abe:2015ibe} and so a co-bimaximal mixing pattern is still well supported by the data. This extension of the Ma model has interesting features: the addition of vector-like fermions increases potential LHC signatures, and the model has the potential for multiple components of cosmological DM~\cite{MaCobimaximal}.  Additionally, any model that utilizes the $A_4$ symmetry could just as easily utilize the double cover known as the binary tetrahedral group, or simply $T^{\prime}$.  Frequently, models that utilize a horizontal symmetry to explain lepton mixing produce a CKM mixing matrix which is diagonal in the symmetry basis (for instance Refs.~\cite{Altarelli:2005yp,Ma:2015fpa,MaCobimaximal,Li:2016nap} are models with horizontal symmetries that are either agnostic about quark mixing or assume a diagonal CKM), however flavor symmetries have been used to explain the Cabibbo angle ~\cite{Ge:2014mpa,Varzielas:2016zuo} and in particular an angle close to the physical Cabibbo angle can be produced by utilizing the doublet representations of $T^{\prime}$~\cite{Frampton:1994rk,Frampton:2007et,Feruglio:2007uu,Frampton:2008bz,Eby:2008uc,Ding:2008rj,Frampton:2009fw,Frampton:2009ut,Frampton:2010uw,Eby:2011ph,Eby:2011aa,Eby:2011qa}, however this flavor symmetry has never been studied before in the context of a scotogenic, or Ma model.  In this paper, a model of radiative lepton and quark masses with a $T^{\prime}$ horizontal symmetry is proposed.  By using the $T^{\prime}$ symmetry and using vector-like quarks to complete the quark mass loops, an angle close to the Cabibbo angle is produced. The model is based on the soft co-bimaximal $A_4$ model from Ref.~\cite{MaCobimaximal}, however the $U(1)_D$ is modified (most notably) where the first two generations of quarks are chiral under this `dark' gauge.  The expanded particle content required to cancel anomalies yields interesting collider signatures, and it is found that the model can support multi-component DM with several distinct cases.
\section{The Model}
The particle content for the current proposal is listed in Table~\ref{Tparticles}, and is based on the recent proposal for cobimaximal $A_4$ neutrino mixing which generates charged lepton and neutrino mass from the loops shown in Fig.~\ref{Flnuloop}~\cite{MaCobimaximal}.  The SM gauge group ($SU(3)_C \times SU(2)_L \times U(1)_Y$) is expanded with a dark gauge $U(1)_D$, and there are several additional discrete symmetries: a dark $Z_2$, a softly broken flavor $Z_2$, and the horizontal symmetry $T^{\prime}$. The `flavor' $Z_2$ forbids tree level masses for the charged leptons, whereas the addition of two dark symmetries ($Z_2$ and $U(1)_D$) as well as a horizontal symmetry allows for the generation of neutrino and lepton masses through the loops from Ref.~\cite{MaCobimaximal} as discussed below.  The non-Abelian discrete symmetry $T^{\prime}$ (also known as the binary tetrahedral group) is the double cover of $A_4$, and has many of the same mulitplication rules as $A_4$ namely~\cite{Ishimori:2010au,Chen:2014tpa}
\begin{equation}
{\bf 3} \otimes {\bf 3} = {\bf 1_0} \oplus {\bf 1_1} \oplus {\bf 1_2} \oplus {\bf 3} \oplus {\bf 3},
\end{equation}
\begin{equation}
{\bf 1_i} \otimes {\bf 1_j} = {\bf 1_{i+j}},
\end{equation}
however $T^{\prime}$ has three doublet representations (${\bf 2_0}\text{, }{\bf 2_1}\text{, }{\bf 2_2}$)~\cite{Ishimori:2010au,Chen:2014tpa}:
\begin{equation}
{\bf 2_i} \otimes {\bf 2_j} = {\bf 1_{i+j}} \oplus {\bf 3},
\end{equation}
\begin{equation}
{\bf 2_i} \otimes {\bf 3} = {\bf 2_i} \oplus {\bf 2_{i+1}} \oplus {\bf 2_{i+2}},
\end{equation}
where $i,j=0,1,2$ $mod$ $3$.  
In addition to the particle content from Ref.~\cite{MaCobimaximal} there is another scalar doublet ($\rho$), a non-trivial $T^{\prime}$ scalar singlet ($\sigma$), up-like ($\mathcal{U}$) and down-like ($\mathcal{D}$) vector-like quarks which are $T^{\prime}$ doublets, and up-like and down-like $T^{\prime}$ singlets ($\mathcal{T}$ and $\mathcal{B}$) which are chiral under $U(1)_D$.  The color-charged particles $\mathcal{U}$ and $\mathcal{D}$ are added to complete the loop in Fig.~\ref{Fqloop} in order to radiatively generate the first two generations of quark masses.  Note that these vector-like quarks have dark charge and are also odd under the additional dark $Z_2$, whereas the particles introduced to cancel anomalies carry the softly broken flavor $Z_2$.  The scalar singlets $\zeta_1$ and $\zeta_2$, which have integer dark charges, are a $T^{\prime}$ triplet and trivial singlet respectively and both receive non-zero vacuum expectation values (VEVs), thus spontaneously breaking $U(1)_D$ to a residual $Z_2$, in which particles with half-integer charges are odd under the residual symmetry and all other particles are even.  Even though the right-handed (RH) quarks carry dark charge while the rest of the SM does not, the proposed model is anomaly free.  The $[SU(2)]^2 U(1)_D$ anomaly is zero since the left-handed (LH) quarks and leptons do not carry dark charge, the $[U(1)_Y]^2 U(1)_D$ anomaly is canceled by the contribution from $\mathcal{T}$ and $\mathcal{B}$:
\begin{eqnarray}
SM:& 2\times 3\left(\frac{2}{3}\right)^2\left(1\right)+ 2 \times 3\left(\frac{-1}{3}\right)^2\left(-1\right) = 2\\
\mathcal{T}/\mathcal{B}:& 3\left(\frac{2}{3}\right)^2\left(-\frac{1}{2}\right)- 3\left(\frac{2}{3}\right)^2\left(\frac{3}{2}\right)
+3\left(-\frac{1}{3}\right)^2\left(\frac{1}{2}\right)- 3\left(\frac{-1}{3}\right)^2\left(-\frac{3}{2}\right) = -2.
\end{eqnarray}
While the $U(1)_Y [U(1)_D]^2$ anomaly is also canceled by the $\mathcal{T}$ and $\mathcal{B}$ contributions:
\begin{eqnarray}
SM:& 2\times 3\left(\frac{2}{3}\right)\left(1\right)^2+ 2 \times 3\left(\frac{-1}{3}\right)\left(-1\right)^2 = 2\\
\mathcal{T}/\mathcal{B}:& 3\left(\frac{2}{3}\right)\left(-\frac{1}{2}\right)^2- 3\left(\frac{2}{3}\right)\left(\frac{3}{2}\right)^2
+3\left(-\frac{1}{3}\right)\left(\frac{1}{2}\right)^2- 3\left(\frac{-1}{3}\right)\left(-\frac{3}{2}\right)^2 = -2.
\end{eqnarray}
That is, the $U(1)_D$ anomalies are canceled between the two light generations of $q_R$ and the chiral anomaly coming from $\mathcal{T}$ and $\mathcal{B}$ in analogy to the $U(1)_{R12}$ model~\cite{Dobrescu:2015asa}. The $[SU(3)_C]^2 U(1)_D$ and mixed gravitational anomalies are canceled between $\mathcal{C}_R-\mathcal{S}_R$ and $\mathcal{T}-\mathcal{B}$ separately. Note that the charge assignment required to cancel the $U(1)_D$ anomaly means that the $\zeta_2$ scalar must gain a VEV in order for these dark chiral vector-like quarks to gain a mass, and since these particles carry color-charge it is important that $v_2$ is relatively large so that these particles can evade existing collider constraints on vector-like quarks. 

The mixing terms between scalars are highly restricted from the dark symmetries and the horizontal symmetry such that the scalar potential terms for the fields that do not receive VEVs are generally of the form
\begin{equation}
V_{\phi_i} = \mu_{i}^2 \phi_i^{\dagger} \phi_i + \lambda_{i} | \phi_i^{\dagger} \phi_i |^2 + \lambda_{H \phi_i} | \Phi^{\dagger} \phi_i |^2 + \lambda_{\zeta_{1} \phi_i} | \phi_i^{\dagger} \zeta_1 |^2 + \lambda_{\zeta_{2} \phi_i} | \phi_i^{\dagger} \zeta_2 |^2,
\end{equation}
along with quartic interaction terms between each scalar field of the form $\lambda_{\phi_i \phi_j} | \phi_i^{\dagger} \phi_j |^2$, where $\phi_i \neq \Phi, \zeta_1, \zeta_2$, and where the structure of $\lambda_{\phi_i \phi_j}$ coefficients is fixed by the $T^{\prime}$ assignment.  The mixing terms between the scalars that do not fit this pattern are $\lambda_l \widetilde{\eta}^{\dagger} \Phi \chi^-$ and $\lambda_q \rho^{\dagger} \Phi \sigma \zeta_1$ which allow for the generation of quark and lepton masses, where the $\lambda_l$ term softly breaks the non-dark $Z_2$, and the $\lambda_{D} \rho^{\dagger} \eta s \zeta_1$ term which allows for potentially interesting DM phenomenology but is not of particular relevance for this study.
\begin{center}
\begin{table}
\caption{Particle Content}
\label{Tparticles}
\centering
\begin{tabular}{| c | c | c | c | c | c | c | c |}
\hline
Particles & $SU(3)_C$ & $SU(2)_L$ & $U(1)_Y$ & $U(1)_D$ & dark $Z_{2}$ & $T^{\prime}$ & $Z_2$ \\
\hline
\hline
SM Particles: & & & & & & & \\
\hline
$(\nu,l)_L$ & 1 & 2 & -1/2 & 0 & + & 3 & + \\
$l_R$ & 1 & 1 & -1 & 0 & + & 3 & - \\
$\Phi$ & 1 & 2 & 1/2 & 0 & + & $1_0$ & + \\
$\mathcal{Q}_L = \begin{pmatrix} (c,s)_L \\ (u,d)_L \end{pmatrix}$ & 3 & 2 & 1/6 & 0 & + & $2_0$ & + \\
$\mathcal{C}_R = (c_R, u_R)$ & 3 & 1 & 2/3 & 1 & + & $2_2$ & + \\
$\mathcal{S}_R = (s_R, d_R)$ & 3 & 1 & -1/3 & -1 & + & $2_1$ & + \\
$(t,b)_L$  & 3 & 2 & 1/6 & 0 & + & $1_0$ & + \\
$t_R$ & 3 & 1 & 2/3 & 0 & + & $1_0$ & + \\
$b_R$ & 3 & 1 & -1/3 & 0 & + & $1_0$ & + \\
\hline
\hline
Fermions: & & & & & & & \\
\hline
$N_{L,R}$ & 1 & 1 & 0 & 1/2 & + & 3 & + \\
$E_{L,R}$ & 1 & 2 & 1/2 & 0 & -  & $1_0$ & + \\
$F_L^0$ & 1 & 1 & 0 & 0 & -  & $1_0$ & + \\
$\mathcal{U}_{L,R}$ & 3 & 1 & 2/3 & 1/2 & - & $2_0$ & + \\
$\mathcal{D}_{L,R}$ & 3 & 1 & -1/3 & -1/2 & - & $2_2$ & + \\
$\mathcal{T}_{L}$ & 3 & 1 & 2/3 & -1/2 & + & $1_0$ & - \\
$\mathcal{T}_{R}$ & 3 & 1 & 2/3 & 3/2 & + & $1_0$ & - \\
$\mathcal{B}_{L}$ & 3 & 1 & -1/3 & 1/2 & + & $1_0$ & - \\
$\mathcal{B}_{R}$ & 3 & 1 & -1/3 & -3/2 & + & $1_0$ & - \\
\hline
\hline
Scalars: & & & & & & & \\
\hline
$\eta$ & 1 & 2 & 1/2 & -1/2 & + & $1_0$ & + \\
$\chi^+$ & 1 & 1 & 1 & -1/2 & + & $1_0$ & - \\
$s$ & 1 & 1 & 0 & 0 & - & 3 & + \\
$\rho$ & 1 & 2 & 1/2 & 1/2 & - & 3 & + \\
$\sigma^0$ & 1 & 1 & 0 & -1/2 & - & $1_2$ & + \\
$\zeta_1^0$ & 1 & 1 & 0 & 1 & + & 3 & + \\
$\zeta_2^0$ & 1 & 1 & 0 & 2 & + & $1_0$ & + \\
\hline 
\end{tabular}
\end{table}
\end{center}
\subsection{Lepton Masses}
\begin{figure}[htb]
\centering
\begin{minipage}{0.45\textwidth}
\centering
\includegraphics[scale=0.8,bb = 150bp 600bp 500bp 800bp,clip]{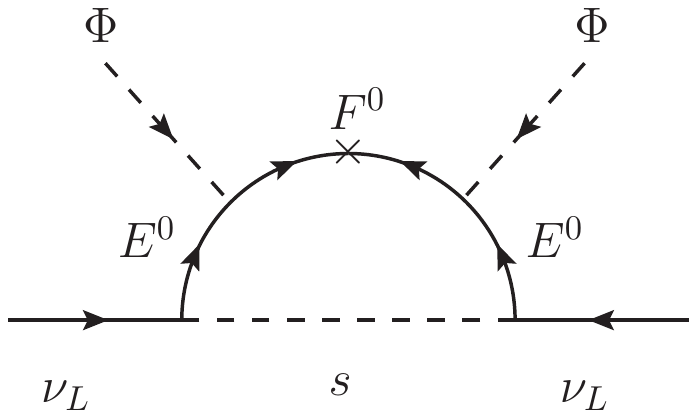}
\end{minipage}
\hspace{10mm}
\begin{minipage}{0.45\textwidth}
\centering
\includegraphics[scale=0.8,bb = 150bp 600bp 400bp 800bp,clip]{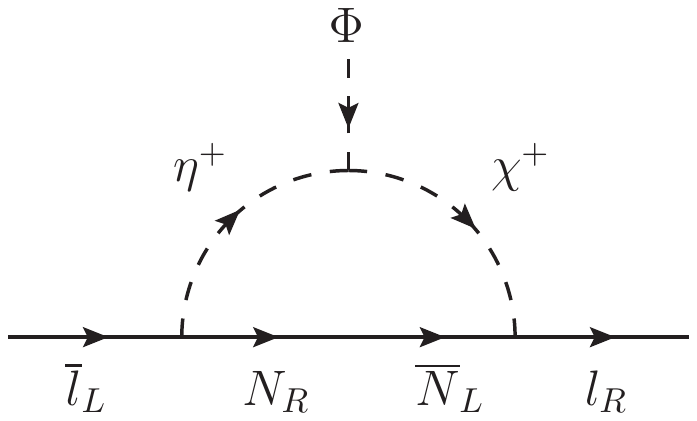}
\end{minipage}
\centering
\caption{One-loop neutrino and lepton masses generated with minimal modification to the mechanism in Ref.~\cite{MaCobimaximal}.}
\label{Flnuloop}
\end{figure}

The charged and neutral lepton masses are generated through the loops shown in Fig.~\ref{Flnuloop}, where the `flavor' $Z_2$ symmetry is softly-broken by the trilinear scalar term.  Note that this is the mechanism used to generate the lepton masses in Ref.~\cite{MaCobimaximal}, however the horizontal symmetry is $T^{\prime}$ instead of $A_4$, and the charge assignment under $U(1)_D$ is chosen to be half-integer values in analogue to Ref.~\cite{Kownacki:2016hpm}, though these changes do not change the predictions for lepton mixing.  Since the leptonic sector only uses $T^{\prime}$ singlets and triplets the mixing pattern is identical to a model utilizing just $A_4$~\cite{MaCobimaximal}, thus the binary tetrahedral model predicts the so-called cobimaximal neutrino mixing pattern ($\theta_{13}\neq0,\theta_{23}=\pi/4,\delta_{CP}=\pm \pi/2$).  
The correct neutrino mass matrix is generated by the soft-breaking of  the $T^{\prime}$ triplet representations to $Z_3$ via the $\overline{N}_L N_R$ masses and to $Z_2$ via $s_1 s_2$ terms as discussed in Ref.~\cite{MaCobimaximal} and resulting masses for the leptons are~\cite{Kownacki:2016hpm}
\begin{equation}
m_l = \frac{f_{lL} f_{lR} \sin (\theta_x) \cos (\theta_x) m_N}{16 \pi^2}\left( F(X_1)-F(X_2)\right),
\end{equation}
where $F(X_i)=X_i \log(X_i)/(X_i-1),$ where $X_{1,2}=m_{x_{1,2}}^2/ m_N^2$ and $m_{x_{1,2}}$ are the scalar masses resulting from the $\lambda_l \widetilde{\eta}^{\dagger} \Phi \chi$ mixing where $\tan(2 \theta_x) =  \frac{2 v \lambda_l}{m_{\chi}^2-m_{\eta}^2}$.  The neutrino masses are given by~\cite{MaCobimaximal}
\begin{equation}
m_{\nu} = \frac{f_{s}^2 m_D^2 m_F}{16 \pi^2 (m_F^2-m_s^2)}\left( G(x_f) - G(x_s)\right),
\end{equation}
where $G(x)=\frac{x}{1-x}+\frac{x^2 \log(x)}{(1-x)^2},$ where $x_F=m_F^2/m_E^2$, $x_s=m_s^2/m_E^2$ and $m_F$ and $m_E$ are the vector-like fermion masses and $m_D$ is the $E-F$ mixing which is assumed to be small~\cite{acnolegma,MaCobimaximal}.  
The importance of this mechanism is that the `clashing' symmetry of the charged and neutral leptons that predicts co-bimaximal mixing ($Z_3\times Z_2$) will generically have arbitrary radiative corrections in the scalar sector if $s_i s_j$ terms break $T^{\prime}$ to $Z_2$ and the scalars generating charged lepton mass break $T^{\prime}$ to $Z_3$, as this forces $Z_3$ breaking counter-terms to be introduced~\cite{MaCobimaximal}.  However by having the dark symmetries ($Z_2$ and $U(1)_D$), and allowing the fermions ($N_i$) to softly break $T^{\prime}$ instead of scalars, these arbitrary corrections are prevented~\cite{MaCobimaximal}.

While this neutrino mixing pattern predicts the maximal $\theta_{23}$, it is possible that slight perturbations or corrections can move this prediction slightly away from this value, though the correlations between mixing angles that are produced from the horizontal symmetry will generally restrict any such deviation similar to the results in Ref.~\cite{acnolegma}.  It is also important to note that scotogenic mechanisms for charged leptons and quarks in general modifies the measured Higgs branching ratios, which has been studied in detail in the context of scotogenic models~\cite{seanma} and have potential important consequences for muon $g-2$~\cite{Nomura:2016ask,Nomura:2016emz}, though such studies are beyond the scope of this work.
\subsection{Quark Masses}
\begin{figure}[htb]
\centering
\begin{minipage}{0.9\textwidth}
\centering
\includegraphics[scale=0.9,bb = 150bp 600bp 420bp 800bp,clip]{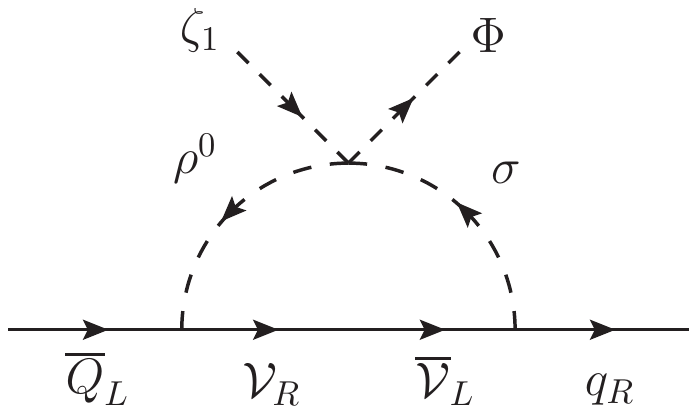}
\end{minipage}
\centering
\caption{One-loop quark masses for the first two generations of quark masses where $\mathcal{V}$ are the corresponding vector-like quarks ($\mathcal{U}$,$\mathcal{D}$).}
\label{Fqloop}
\end{figure}
\begin{figure}[htb]
\centering
\centering
\includegraphics[scale=1]{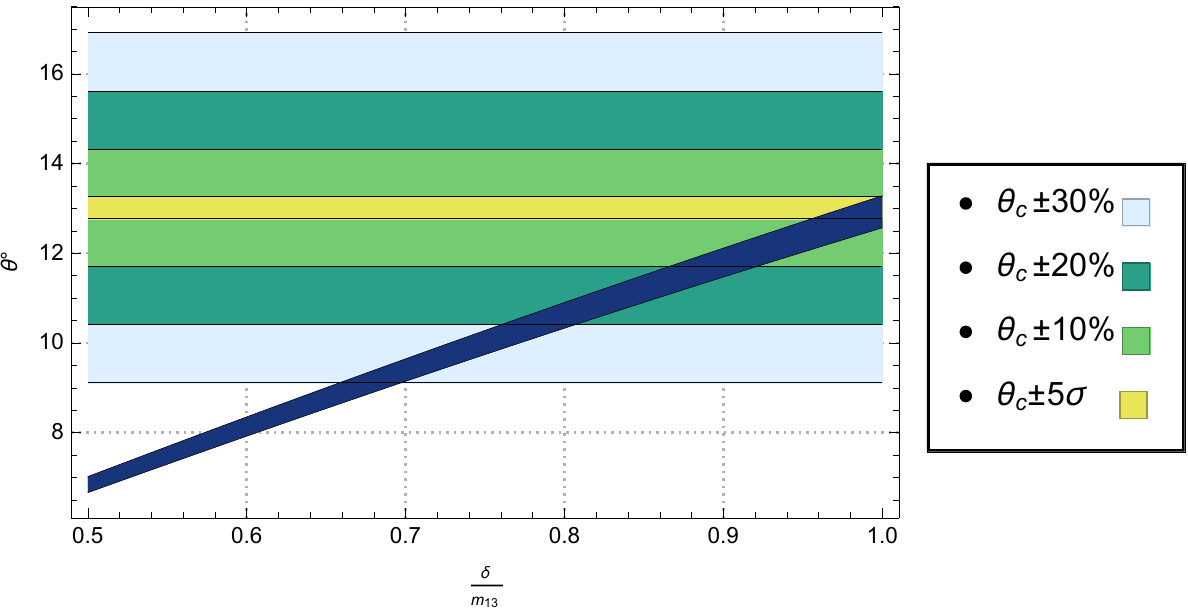}
\centering
\caption{The first-order value of CKM mixing angle $\frac{| V_{us} |}{|V_{ud}|}$ in degrees as a function of the $T^{\prime}$ soft-breaking terms $\delta/m_{13}$.  Banded region represents the physical Cabibbo angle ($\theta_c$) $\pm 30\%,20\%,10\%,$ and $5 \sigma$.}
\label{Cabibbo}
\end{figure}

The loop for the generation of the $u,d,c,$ and $s$ quark masses is shown in Fig.~\ref{Fqloop}, where the loop is completed by the $T^{\prime}$ triplet scalar doublet $\rho$ and the non-trivial $T^{\prime}$ singlet scalar $\sigma^0$, and the $T^{\prime}$ symmetry is softly-broken by the vector-like quark masses which are generated by dimension three terms.  The mass matrix for $\rho$ is given by
\begin{equation}
\mathcal{M}_{\rho} =\begin{pmatrix}
 A & B & B \\ B & A & B \\ B & B & A
 \end{pmatrix},
\end{equation}
where A is a combination of $v^2$,$v_1^2$, and $v_2^2$ and B is proportional to just $v_1^2$.  This mass matrix is exactly diagonalized by the tribimaximal mixing matrix:
\begin{equation}
U_{TB} = \begin{pmatrix}
\sqrt{2/3} & 1/\sqrt{3} & 0 \\
-1/\sqrt{6} & 1/\sqrt{3} & -1/\sqrt{2} \\
-1/\sqrt{6} & 1/\sqrt{3} & 1/\sqrt{2} 
\end{pmatrix},
\end{equation}
since B only depends on $v_1^2$ then it is a reasonable assumption that if $v_2 > v_1$ that the resulting masses for $\rho$ are nearly degenerate.  
The quark mass loop is completed by the $\lambda_{q} \Phi^{\dagger} \rho \sigma \zeta^{\star}_1$ term, so the mixing is relatively small.  Given the assumption that the $\mathcal{M}_{\rho}$ masses are nearly degenerate after rotating with the tribimaximal matrix, then the mass matrix that spans the $\rho-\sigma$ states, $\mathcal{M}_{\rho \sigma}$, is of the form
\begin{equation}
\mathcal{M}_{\rho \sigma} = \begin{pmatrix}
m_{\rho}^2 & \lambda_{q} v_{\Phi} v_{\zeta} \\
\lambda_{q} v_{\Phi} v_{\zeta} & m_{\sigma}^2 
\end{pmatrix},
\end{equation}
where the resulting mass states of $\mathcal{M}_{\rho \sigma}$ are labeled as $y_{1,2}$.
The quark mass matrix is thus given by 
\begin{equation}
\mathcal{M}_{q} = \frac{f_{qL} f_{qR} \sin(\theta_{y}) \cos(\theta_{y})}{32 \pi^2} \mathcal{I}_q,
\end{equation}
where $\tan(2 \theta_{y})= \frac{2 \lambda_q v v_{1}}{m_{\sigma}^2-m_{\rho}^2}$ and $\mathcal{I}_q$ is a 2x2 matrix where the flavor structure and the loop calculations has been taken into account. Specifically, for the up-like quarks $\mathcal{I}_q$ is of the form
\begin{equation}
\begin{pmatrix}
-(\mathcal{F}[X_1] \cos(\theta_{\mathcal{V}})^2+\mathcal{F}[X_2]\sin{(\theta_{\mathcal{V}})}^2) & (\mathcal{F}[X_1]-\mathcal{F}[X_2])\sin{(2 \theta_{\mathcal{V}})}\\
- (\mathcal{F}[X_1]-\mathcal{F}[X_2])\sin{(2 \theta_{\mathcal{V}})}  & \mathcal{F}[X_1] \sin{(\theta_{\mathcal{V}})}^2+\mathcal{F}[X_2]\cos{(\theta_{\mathcal{V}})}^2
\end{pmatrix},
\end{equation}
with $\mathcal{F}[X_i]= F[X_{i1}] - F[X_{i2}]$, and $F[X_{ij}]= m_{\mathcal{V}_i}X_{ij} \log{(X_{ij})}/(X_{ij}-1)$,
$X_{ij}=m_{y_j}^2/m_{\mathcal{V}_i}^2$, and $\theta_{\mathcal{V}}$ is the mixing angle that diagonalizes $\mathcal{M}_{\mathcal{V}}$ (the vector-like quark mass matrix),
and for the down-like quarks $\mathcal{I}_q$ is
\begin{equation}
\sqrt{2}
\begin{pmatrix}
0 & \mathcal{F}[X_{1}] \cos^2(\theta_{\mathcal{V}}) +\mathcal{F}[X_2] \sin^2(\theta_{\mathcal{V}})\\
0  & (\mathcal{F}[X_{1}]  - \mathcal{F}[X_2]) \cos(\theta_{\mathcal{V}}) \sin(\theta_{\mathcal{V}})
\end{pmatrix},
\end{equation}
where the miss-match is because of the different transformations of the vector-like quarks $\mathcal{U}$ and $\mathcal{D}$ under the horizontal symmetry ($T^{\prime}$).
  The quark mass matrix squared,
 $\mid \mathcal{M} \mid^2$, is diagonalized by a 2x2 rotation matrix where the angle is a function of $\theta_{\mathcal{V}}$.  The resulting CKM matrix is the
miss-match between the up-like and down-like sectors.  The $T^{\prime}$ symmetry is softly broken by dimension three terms of $\overline{\mathcal{V}}_L \mathcal{V}_R$ of the form
\begin{equation}
\mathcal{M}_{\mathcal{U},\mathcal{D}}=
\begin{pmatrix}
\frac{m_{11} + m_{33} + 2 m_{13}}{4}& \frac{i (m_{11} - m_{33})}{4}\\
\frac{-i (m_{11} - m_{33})}{4} & \frac{m_{11} + m_{33} - 2 m_{13}}{4}
\end{pmatrix},
\label{EqSoftBreak}
\end{equation}
which is protected by a residual $Z_4$ symmetry where 
\begin{equation}
\mathcal{V}_1= (\mathcal{V}_{1_0}+\mathcal{V}_{1_3})/2,
\end{equation}
and 
\begin{equation}
\mathcal{V}_2= -i(\mathcal{V}_{1_1}-\mathcal{V}_{1_3})/2,
\end{equation}
where
$\mathcal{V}_{1_1}$, and $\mathcal{V}_{1_3}$ are the $Z_4$ flavor states transforming as non-trivial singlets ${\bf 1_1}$ and ${\bf 1_3}$ respectively (where ${\bf 1_0}$ is the trivial singlet) and where $\overline{\mathcal{V}}_{1^i} \mathcal{V}_{1^j}=m_{ij}$ and $m_{13}=m_{31}$.  Note that this basis where Eq.~\ref{EqSoftBreak} has complex off-diagonals can be rotated in such a way where in the $T^{\prime}$ limit the mass matrix is given by equal diagonal values.  In this basis, the soft-breaking matrix can be parameterized assuming $m_{33}=m_{11}+\delta$ which yields
\begin{equation}
\begin{pmatrix}
2 m_{11} + \delta + 2 m_{13} & \delta \\
\delta & 2 m_{11} + \delta - 2 m_{13} 
\end{pmatrix},
\end{equation} 
if $m_{13} \ll 1$ and $\delta \propto m_{13}$ then the deviation from the $T^{\prime}$ symmetric mass matrix is small.  It is convenient to parameterize the soft-breaking matrix in terms of the mass eigenstate $m_{11}$ such that $m_{13} = \epsilon m_{11}$.  The rotation that diagonalizes the vector-like quarks is the same for up-like and down-like $\mathcal{V}'s$, but the differing textures of $\mathcal{M}_q$ produce different dependence on $\theta_{\mathcal{V}}$ for the $V_L^q$ that diagonalizes $\mid \mathcal{M} \mid^2$.  For the up-like sector the angle $\theta_{\mathcal{U}}$ is simply  
$\tan(2 \theta_{\mathcal{U}}) = \delta/(2 m_{13})$, whereas for the down-like sector is approximately $\tan(2 \theta_{\mathcal{D}}) = \delta \epsilon /(2 m_{13})$.  The resulting Cabibbo angle as a function of $\delta/m_{13}$, and for various $\epsilon \leq 1/2$, are plotted in Fig.~\ref{Cabibbo}.  While the exact Cabibbo angle can only be fit for very specific choices of the soft-breaking terms, a variation on the order of 30$\%$ from the physical value can be fit in a much wider parameter space.  
 It is important to note that in order to fit all of the parameters of the physical $V_{CKM}$ it is well known that a perturbation to this texture from running effects or some higher level loop contributions must exist~\cite{Varzielas:2016zuo}, or even simply additional $T^{\prime}$ doublets at a higher mass scale.  However if the predicted mixing angle, $\theta_c$, is within 15 percent of the physical Cabibbo angle then the relative difference between $\theta_c$ and the physical Cabibbo angle is the same size as the next largest mixing angle in the CKM; in comparison to the implementation of quark mixing in Ref.~\cite{MaCobimaximal}, the CKM after soft-breaking of the horizontal symmetry is still approximately diagonal and thus the deviation from the physical values is quantitatively smaller in this Binary Tetrahedral model.  

\begin{figure}[htb]
\centering
\begin{minipage}{0.9\textwidth}
\centering
\includegraphics[scale=0.9,bb = 150bp 600bp 420bp 800bp,clip]{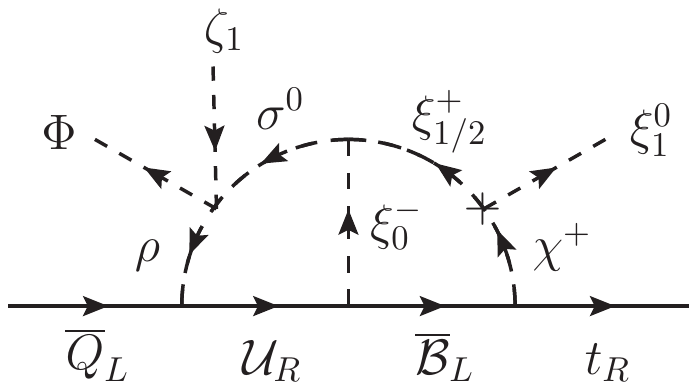}
\end{minipage}
\centering
\caption{Example of a two-loop extension to generate mixing between the first two generations of quarks and the top quark, where $\xi_0^{\pm} \sim (0,{\bf 2_i},-,-)$, $\xi_{1/2}^{\pm} \sim (1/2,{\bf 2_j},+,-)$, and $\xi^0_1 \sim (1,{\bf 2_k},+,+)$ under ($U(1)_D,T^{\prime}$, dark $Z_2$, $Z_2$), and $\xi^0_1$ receives a non-zero VEV.}
\label{TwoLoop}
\end{figure}

For a proof of concept of how this model could accomodate a fully realistic CKM, consider the two-loop mixing of the first generations with $t$ as shown in Fig.~\ref{TwoLoop}, where $\xi_0^{\pm} \sim (0,{\bf 2_i},-,-)$, $\xi_{1/2}^{\pm} \sim (1/2,{\bf 2_j},+,-)$, and $\xi^0_1 \sim (1,{\bf 2_k},+,+)$ under ($U(1)_D,T^{\prime}$,dark $Z_2$,$Z_2$), and $\xi^0_1$ receives a non-zero VEV and each new scalar is a singlet under $SU(2)$.  This adds additional complications to the scalar sector, however $\xi_1^0$ is a $T^{\prime}$ doublet and only couples to the other scalars via quartic terms, and so these additional scalars will not contribute to the Higgs mass directly if each component of the $T^{\prime}$ doublet receive the same VEV.  This consequence is easiest to determine in the Ma$-$Rajasekaran basis from Ref.~\cite{Chen:2014tpa}; in this basis $\overline{\xi_1^0} \xi_1^0 = \frac{1}{\sqrt{2}} (\overline{\xi^0_{11}} \xi^0_{12} -  \overline{\xi^0_{12}} \xi^0_{11})$, where $\xi_{1i}^0$ is the $i$th component of the $T^{\prime}$ doublet and where $\langle \xi_{11}^0 \rangle = \langle \xi_{12}^0 \rangle \neq 0$.  This ultimately leads to only trilinear terms between the physical degrees of freedom of $\xi_{1i}^0$ and $\Phi$ (after taking into account additional scalar mixing terms between $\xi^0_1$ and $\zeta_1$) but has no contributions to the SM-like Higgs boson mass, regardless of which of the distinct $T^{\prime}$ doublet representations are chosen for $\xi$, and so there are no clearly problematic contributions to flavor physics from these new scalars.  There is non-trivial $\xi_1^0-\zeta_1$ mixing as the multiplication rules of $T^{\prime}$ (${\bf 2_i} \otimes {\bf 2_j} = {\bf 3}$) generate the appropriate terms in the scalar potential which are not eliminated after the new constraint equations are taken into account. In the mass matrix spanned by the scalar degrees of freedom, these mixing terms will contribute to $\zeta_1-\xi_1^0-\zeta_2$ and $Im(\zeta_1) - Im(\xi_1^0)$ mixing, but the details of this mixing and the proper prediction of the CKM matrix depend on the exact choice of $T^{\prime}$ representations chosen for the $\xi$ scalars.  These additional scalars will alter the mass matrices for the first two generations of quarks in order to generate the proper CKM, thus the physical psuedoscalar couplings and masses should be correlated to the perturbation of the one-loop quark mixing angles.  These scalars also allow new interactions for the vector-like quarks, however if $m_{\xi}$ is large these new decay channels may be suppressed at the LHC.  In addition, these new scalars could be treated as mediators for the dark sector if the lightest scalar resulting from the $\sigma^0-\rho$ mixing is the DM candidate for the sector that carries charge under the additional $U(1)$ gauge group.

\subsection{Scalar Sector}
There are three scalars with integer charges under the $U(1)_D$ gauge symmetry, where $\Phi$ breaks $SU(2)_L \times U(1)_Y$ just as the Higgs field in the SM, where the VEVs of the scalars charged under $U(1)_D$ ($\langle \zeta_1i \rangle = v_{1i}$ and $\langle \zeta_2 \rangle = v_2$) break the dark gauge to a residual $Z_2$ dark parity, where particles with half-integer charges have odd parity and all others have even parity.  The scalar potential relevant to the symmetry breaking is
\begin{equation}
\begin{split}
\displaystyle
V &= \mu_{H}^2 \Phi^{\dagger} \Phi +\frac{\lambda_H}{2} (\Phi^{\dagger} \Phi)^2 + \mu_1^2 \sum_i \zeta_{1i}^{\star} \zeta_{1i} + \frac{\lambda_{1_0}}{2} (\zeta^{\star}_{1i} \zeta_{1i})^2  \\
 &+ \lambda_{1_1}
(\sum_i \omega^{2(i-1)} \zeta_{1i}^{\star} \zeta_{1i})(\sum_i \omega^{i-1} \zeta_{1i}^{\star} \zeta_{1i}) + \lambda_{13}
(\sum_i \zeta^{\star}_{1(i+2)} \zeta_{1i} \zeta^{\star}_{1i} \zeta_{1(i+2)})\\
&+ \frac{\lambda_{13^{\prime}}}{2} ((\zeta^{\star}_{2} \zeta_{3})^2
+(\zeta^{\star}_{3} \zeta_{1})^2+(\zeta^{\star}_{1} \zeta_{2})^2 + H.C.)\\
&+ \mu_2^2 \zeta_2^{\star} \zeta_2 + \frac{\lambda_2}{2} (\zeta_{2}^{\star} \zeta_2)^2 \frac{\mu_{12}}{2} \sum_i ( \zeta_{1i} \zeta_{1i} \zeta_2^{\star} + H.C.)\\
&+ \sum_i \lambda_{H1} \Phi^{\dagger} \Phi \zeta_{1i}^{\star} \zeta_{1i} + \lambda_{H2} \Phi^{\dagger} \Phi \zeta_2^{\star} \zeta_2
\end{split}
\label{SSBV}
\end{equation}
where $\mu_{12}$ can be taken to be by rotating the relative phase between $\zeta_{1i}$ and $\zeta_2$, and $\omega = e^{i \frac{2 \pi}{3}}$ (note that $\omega^k$ terms are modulo 3).  If the $\Phi - \zeta_{1}$ and $\zeta_2 - \zeta_1$ mixing is ignored, then minimizing the potential for $\zeta_{11}$ yields:
\begin{equation}
\begin{split}
\frac{1}{\zeta_{11}^{\star}}\frac{\partial{V}}{\partial{\zeta_{11}}} &= 
\mu_1^2  + \lambda_{1_0} \displaystyle \Sigma_i(v_{1i}^2) + \lambda_{1_1} (v_{11}^2 + \omega^2 v_{12}^2 + \omega v_{13}^2)(v_{11}^2 + \omega v_{12}^2 + \omega^2 v_{13}^2) \\
&+ \lambda_{13} v_{11} (v_{12}^2 + v_{13}^2) + \lambda_{13^{\prime}} v_{11}(v_{12}^2 + v_{13}^2)
\end{split},
\end{equation}
with similar terms for $\zeta_{12}$ and $\zeta_{13}$. These constraints are met given
\begin{equation} 
v_{11}=v_{12}=v_{13}= v_{1}=\sqrt{\frac{-\mu_1}{3\lambda_{1_0}+2 \lambda_{13} + 2 \lambda_{13^{\prime}}}},
\label{z1vevapprox}
\end{equation}
and since both $\Phi$ and $\zeta_2$ are trivial $T^{\prime}$ singlets the corrections to Eq.~\ref{z1vevapprox} will be equal for all $\zeta_{1i}$, so this minimization condition can be satisfied even with the mixing terms if all $\zeta_{1i}$ VEVs are equal.  With $v_{11}=v_{12}=v_{13}= v_{1}$ and $\zeta_{11}=\zeta_{12}=\zeta_{13}=\zeta_{1}$ the full constraint equations are
\begin{eqnarray}
0 &=& \mu_{H}^2 + \lambda_{H} v^2 + 3 \lambda_{H1} v_1^2+ \lambda_{H2} v_2^2\\
0 &=& \mu_1^2 + 3 \lambda_{1_0} v_1^2+2(\lambda_{13} +  \lambda_{13^{\prime}})v_1^2+\mu_{12} v_2 + \lambda_{H1} v^2 + \lambda_{12} v_2^2\\
0 &=& \mu_2^2 + \lambda_2 v_2^2 + \frac{3}{2} \mu_{12} \frac{v_1^2}{v_2} +  \lambda_{H2} v^2 + 3 \lambda_{12}v_1^2.
\end{eqnarray}
The resulting mass matrix for $h$,$\sqrt{2} Re(\zeta_{1})$, and $\sqrt{2} Re(\zeta_{2})$ is 
\begin{equation}
\begin{pmatrix}
2 \lambda_H v^2  & 6  \lambda_{H1} v v_1  & 2  \lambda_{H2} v v_2 \\ 
6 \lambda_{H1} v v_1  & 6 (\lambda_1 + 2(\lambda_{13} +  \lambda_{13^{\prime}})) v_1^2 &  3 (2 \lambda_2 v_2 +\mu_{12}) v_1 \\
2 \lambda_{H2} v v_2  & 3 (2 \lambda_2 v_2 +\mu_{12})  v_1 & 2 \lambda_2 v_2^2  - \frac{3}{2} \mu_{12} \frac{v_1^2}{v_2}
\end{pmatrix}.
\end{equation}
In general, there is mixing between the $\zeta_1$ and $\zeta_2$ scalars, however for simplicity we take the term $3(\lambda_2 v_2 + \mu_{12})v_1$ to be negligible, and $v^2 \ll v_{1,2}^2$ so that $\phi^{\pm}$, and $\sqrt{2} Im(\phi^0)$ become the longitudinal components of the $W^{\pm}$ and Z gauge bosons, and $h$ is the 125 GeV SM Higgs (note that since $v_2$ is responsible for the $\mathcal{T}_{R,L}$ and $\mathcal{B}_{R,L}$ masses that the assumption that $v_2^2 \gg v^2$ is necessary for the masses of $\mathcal{T}$ and $\mathcal{B}$ to be much larger than the SM quarks).  The mass matrix for
$\sqrt{2} Im(\zeta_{1,2})$ is
\begin{equation}
\begin{pmatrix}
-6 \mu_{12} v_2   & 3 \mu_{12} v_1 \\
 3 \mu_{12} v_1 &  -\frac{3 \mu_{12} v_1^2}{2 v_2}
\end{pmatrix},
\end{equation}
which is diagonal for the linear combinations of $\sqrt{2}(v_1 Im(\zeta_{1}) + 2 v_2 Im(\zeta_{2}))/\sqrt{v_1^2+4v_2^2}$, which is massless and becomes the longitudinal mode for the $Z^{\prime}$ boson, and $\sqrt{2}(v_1 Im(\zeta_{2}) - 2 v_2 Im(\zeta_{1}))/\sqrt{v_1^2+4v_2^2}$ which becomes the massive pseudoscalar particle $A$.  The physical scalar masses are thus
\begin{equation}
m_{h}^2 \approx  2 v^2 \left( \lambda_H - 
\frac{3 \lambda_{H1}^2}{2(\lambda_{13} + \lambda_{13^{\prime}})}- 2 \frac{ v_2^2 \lambda_{H2}^2}{\lambda_2 v_2^2-\frac{3}{2} \mu_{12} \frac{v_1^2}{v_2}} \right),
\end{equation}
\begin{equation}
m_{\zeta_{1R}}^2 \approx 6 v_1^2 (\lambda_{1_0}+2(\lambda_{13} + \lambda_{13^{\prime}}))\text{, }
m_{\zeta_{2R}}^2 \approx v_2^2(\lambda_2 - \frac{3}{2} \mu_{12} \frac{v_1^2}{v_2^3})\text{, }m_{A}^2\approx-\frac{3 \mu_{12} (v_1^2 + 4 v_2^2)}{2 v_2}.
\end{equation}
Many of the fermions with $U(1)_D$ charge are chiral, and have no Yukawa couplings at tree-level to $\zeta_{1i}$ (or $\zeta_{2}$) with the exception of $\mathcal{T}$ and $\mathcal{B}$, which are given by
\begin{equation}
m_{\mathcal{T}} = v_2 y_{\mathcal{T}} \text{, } m_{\mathcal{B}}= v_2 y_{\mathcal{B}},
\end{equation}
thus for Yukawa couplings on order of one, and $v_2 \gg v$, these vector-like quark masses can be large.

The mass eigenvalues from $\rho - \sigma$ and $\eta - \chi$ are given by
\begin{eqnarray}
m_{y^0_{1,2}}&=& \frac{1}{2} \left[3 m_{\rho}^2 +m_{\sigma}^2 + \cos ( 2 \theta_y ) (\pm m_{\rho}^2 \mp m_{\sigma}^2) \pm 6 v v_1 \lambda_q \sin (2 \theta_y)\right] \\
m_{y^{\pm}} &=& \frac{1}{2} m_{\rho}^2 \\
m_{x^{\pm}_{1,2}}^2 &=& \frac{1}{2} \left[ m_{\eta}^2 + m_{\chi}^2 + \cos(2 \theta_x)(\pm m_{\eta}^2 \mp m_{\chi}^2) + 2 v \lambda_l \sin(2 \theta_x) \right] \\
m_{x^0} & = & \frac{1}{2} m_{\eta}^2,
\end{eqnarray}
where $\tan(2 \theta_x) =  \frac{ 2 v \lambda_l }{m_{\chi}^2-m_{\eta}^2}$, $\tan (2 \theta_y) = \frac{2 v v_1 \lambda_q}{m_{\sigma}^2-m_{\rho}^2}$, and there are three copies of $y^{\pm}$ that are nearly degenerate. 
\subsection{Dark Gauge Sector}
The relevant Lagrangian terms for the $Z^{\prime}$ gauge boson are
\begin{equation}
\mathcal{L} \supset -\frac{1}{4} Z^{\prime \mu \nu} Z^{\prime}_{\mu \nu} -\frac{\sin(\kappa)}{2} Z^{\prime \mu \nu} B_{\mu \nu} + (D_{\mu} \zeta_{1i})^{\dagger}(D^{\mu} \zeta_{1i}) +
(D_{\mu} \zeta_2)^{\dagger}(D^{\mu} \zeta_2) -V,
\end{equation}
where $D_{\mu}$ is the covariant derivative ($D_{\mu} = \partial_{\mu} + i q_{D} g_{\zeta} Z^{\prime}_{\mu}$) and $V$ is the scalar potential from Eq.~\ref{SSBV}.  Expanding $\zeta_{1i}$ and $\zeta_{2}$ under the assumptions in the previous section, yields
\begin{equation}
m_{Z^{\prime}}^2 \approx g_{\zeta}^2 ( 3 v_1^2 + 4 v_2^2),
\end{equation}
and where the $\zeta Z^{\prime}Z^{\prime}$ term becomes
\begin{equation}
2 g_{\zeta}^2(3 Re(\zeta_1)  v_1+4 Re(\zeta_2) v_2)Z^{\prime}_{\mu} Z^{\prime \mu},
\end{equation}
and the $\zeta\zeta Z^{\prime}Z^{\prime}$ becomes
\begin{equation}
g_{\zeta}^2 (3 Re(\zeta_1)^2+4 Re(\zeta_2)^2)Z_{\mu}^{\prime} Z^{\prime \mu}
\end{equation}
The structure of the $U(1)_D$ charges makes this model very similar to leptophobic $Z^{\prime}$ models since the only tree-level coupling to the SM is to quarks.  There is no tree-level kinetic mixing term, however such a term can arise at the loop-level due to the coupling to quarks which generates the $\kappa$ term, and so there is some non-zero coupling to leptons.  This kinetic mixing can have an important impact on the dark sector constraints~\cite{Bell:2016fqf,Duerr:2016tmh}, though for the purposes of this study the mixing is taken to be negligible.

There are some notable differences in this model compared to many common $Z^{\prime}$ models, namely the top and bottom quarks do not directly couple to the $Z^{\prime}$ so collider searches of the form $pp \rightarrow Z^{\prime} \rightarrow t\overline{t}$ do not apply.  Mixing contraints involving the $b$ quark do not occur at tree-level, and the first two up-like quarks carry opposite charges to the first two down-like quarks.  There are additional considerations as leptophobic $Z^{\prime}$ models generically produce FCNCs in the RH quark decays~\cite{Barger:2003hg,GonzalezAlonso:2012jb} which can be heavily constraining.  However, $K^0 - \overline{K}^0$ and $B^0-\overline{B}^0$ are less important since the LH quarks do not carry dark charge~\cite{Baek:2006bv,Leroux:2001fx}. Previous studies of $Z^{\prime}$ with only RH coupling to quarks have been carried out, and indicate that a $Z^{\prime}$ with a mass on order of 1 TeV is still viable for certain mixing constraints, particularly in a case with a $U(2)$ flavor symmetry~\cite{Buras:2012jb} which can be accomodated by the $T^{\prime}$ horizontal symmetry~\cite{Aranda:1999kc}.  There are also additional constraints on leptophobic models from the early LHC searches (cf Ref.~\cite{Alves:2013tqa,Fairbairn:2016iuf}), and from the 13 TeV run~\cite{ATLAS-CONF-2016-030,ATLAS-CONF-2016-069,ATLAS-CONF-2016-029}, however there is still a viable range of parameter space for $Z^{\prime}$ models with a $g_{\zeta} \approx 0.1$ and masses between $1\text{ TeV } < m_{Z^{\prime}} < 1.5$ TeV.  Another interesting possibility is if the coupling constant $g_{\zeta} \ll 1$, which allows for much lighter $Z^{\prime}$'s (albeit very weakly interacting)~\cite{Gondolo:2011eq}.  The addition of vector-like quarks that couple to the $Z^{\prime}$ are another source of FCNC, however mixing of $\mathcal{T}-t$ and $\mathcal{B}-b$ are forbidden at tree-level from the flavor $Z_2$ assignment.  If neither the horizontal or the $Z_2$ flavor symmetry are broken by any hard terms, then the RH coupling to the $Z$ and $Z^{\prime}$ are fixed, and so it is reasonable to assume the RH quark mixing to be exactly diagonal at first order which eliminates FCNCs in the gauge sector.  Some amount of FCNC are unavoidable in the scalar sector but these are severely restricted by the additional symmetries, that is the gauge and $T^{\prime}$ assignments prevent tree-level FCNCs and the soft-breaking terms of $T^{\prime}$ are in the fermion sector.  However, there are still contributions to the FCNCs from box and penguin diagrams.  For instance, the largest source of FCNCs in the box diagrams are from the charged scalar $\rho$ and the vector-like quarks $\mathcal{U}$ and $\mathcal{D}$, which generate $K^0-\overline{K^0}$ mixing through the dimension six $(\bar{s}_R d_L)(\bar{s_L} d_R)$ operator which can be estimated as
\begin{equation}
\frac{c}{\Lambda^2} \approx \frac{\delta^2}{32 \pi^2 ( 4 m_{13}^2 +\delta^2)} \frac{f_{qL}^2 f_{qR}^2 (m_{y^{\prime}}^2-m_{y}^2)}{(m_{\mathcal{V}_i}^2-m_{y^{\prime}}^2)(m_{\mathcal{V}_j}^2-m_{y}^2)} \approx \frac{10^{-7} \Delta m_{y y^{\prime}}^2 }{\Delta m_{i y^{\prime}}^2 \Delta m_{jy}^2},
\end{equation}
where $\delta$ and $m_{13}$ are from the $\theta_{\mathcal{V}}$ mixing angle, $m_{y}$ ($m_{y^{\prime}}$) are the masses of the $\rho - \sigma$ mixing and $m_{\mathcal{V}_{i,j}}$ are components of the appropriate vector-like $T^{\prime}$ doublets, with the assumption that the couplings $f_{qL,R} \sim 0.1$.  The $\Delta m_{y y^{\prime}}^2$ term is constrained by the oblique parameters and other electroweak precision tests, whereas $\Delta m_{i y}^2$ can be quite large, and so the overall contribution to FCNCs will in general be below existing upper-limits~\cite{Isidori:2014rba}.  This suppression is a general feature of the box diagrams in this model, and so similar expressions can apply to different dimension six FCNC operators.  There are also penguin diagram contributions, though the loop structure is similar to the one-loop diagrams that generate the up and down-like quark masses from Fig.~\ref{Fqloop}.  Since these loops involve the vector-like quarks and the doublets $\rho_i$, the contribution to the leptonic FCNCs decays, such as $B_s^0 \rightarrow l^{\pm} l^{\mp}$, will primarily arise from $Z^{\prime}$ kinetic mixing term which is small.  That is, while $\eta-\chi$ mixing can introduce some leptonic FCNC terms, there is no tree-level $\rho-\eta$ mixing which means there is no direct connection between the additional scalars that couple to quarks and the scalars that couple to leptons.
\subsection{Dark Matter}
\begin{figure}[htb]
\centering
\hspace{-5mm}
\begin{minipage}{0.3\textwidth}
\centering
\hspace{-20mm}
\includegraphics[scale=0.675,bb = 150bp 600bp 400bp 800bp,clip]{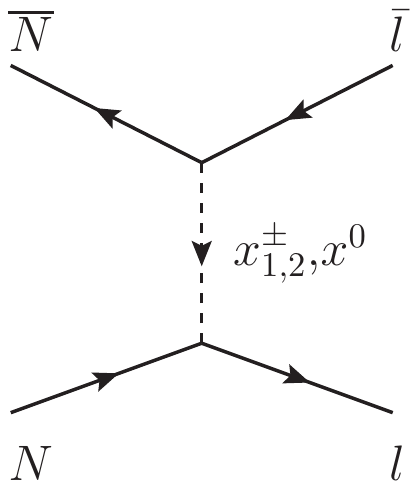}
\end{minipage}
\hspace{5mm}
\begin{minipage}{0.3\textwidth}
\centering
\hspace{-15.5mm}
\includegraphics[scale=0.675,bb = 150bp 600bp 400bp 800bp,clip]{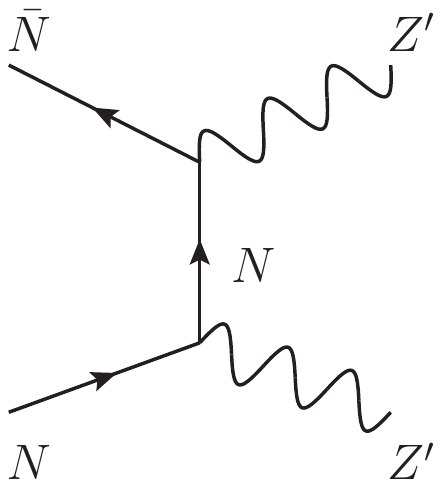}
\end{minipage}
\begin{minipage}{0.3\textwidth}
\centering
\hspace{-12mm}
\includegraphics[scale=0.65,bb = 150bp 600bp 400bp 800bp,clip]{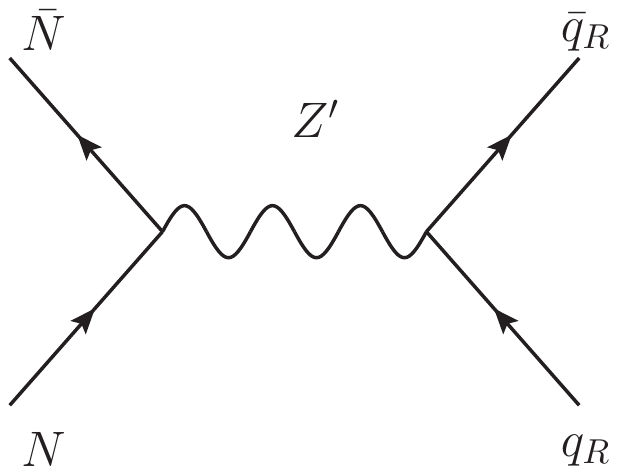}
\end{minipage}

\centering
\caption{Relevant diagrams for $N N$ annihilation contribution to $\langle \sigma v \rangle$.}
\label{NNFeynRelic}
\end{figure}

\begin{figure}[htb]
\centering
\begin{minipage}{0.45\textwidth}
\centering
\hspace{-10mm}
\includegraphics[scale=0.65,bb = 150bp 600bp 400bp 800bp,clip]{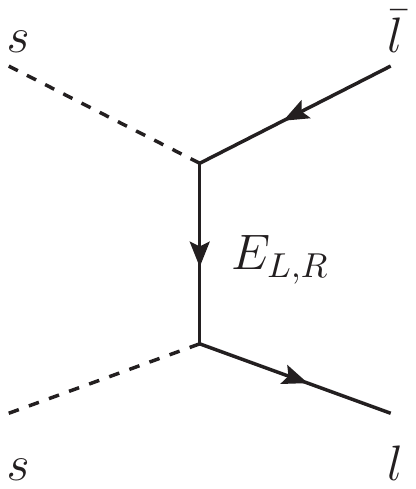}
\end{minipage}
\begin{minipage}{0.45\textwidth}
\centering
\hspace{-10mm}
\includegraphics[scale=0.65,bb = 150bp 600bp 500bp 800bp,clip]{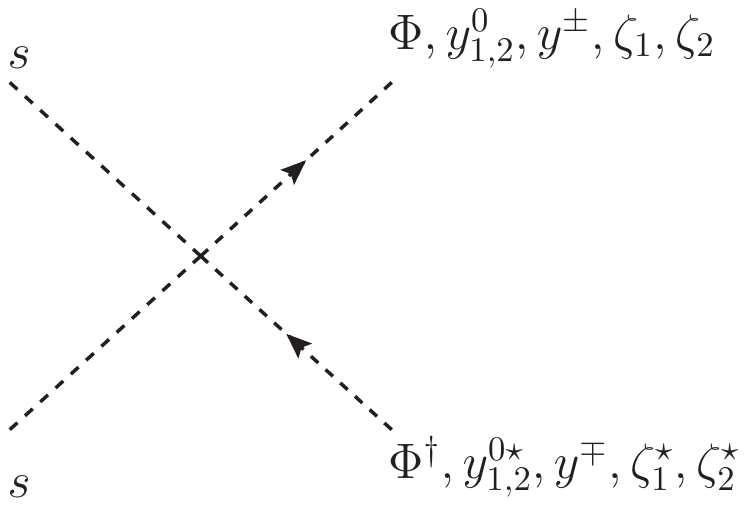}
\end{minipage}
\centering
\caption{Relevant diagrams for $s s$ annihilation contribution to $\langle \sigma v \rangle$.}
\label{SSFeynRelic}
\end{figure}

The dark sector is potentially very complicated: in the dark-charged sector the DM candidate is the lightest mass state of any of the neutral scalars ($\eta^0$, $\rho^0_i$, and $\sigma^0$) or the lightest neutral fermions $N_i$, and for the dark-parity-only sector it is potentially the lightest neutral fermion ($E^0$ of $F^0$) or $s_i$.  However, there are numerous constraints on scalar DM particles that interact with the electroweak (EW) gauge bosons, so to avoid these we set the mass spectrum such that $N_1$ and $s_1$ are the only DM candidates.  The main interactions relevant for the relic density are show in Fig.~\ref{NNFeynRelic} for the fermionic DM candidate, and in Fig.~\ref{SSFeynRelic} for the scalar DM candidates. Even in this framework however, there are several major scenarios: either N or s are the bulk of the cosmological DM or each species significantly contributes to the cosmological DM.  Before any further assumptions of masses are made, the relevant SM-DM couplings are given in the mass basis by
\begin{equation}
f^{\prime}_{lL} \left[ \overline{e}_L N_{1R}(\cos(\theta_x) x_1^- - \sin(\theta_x) x_2^-) + \overline{\nu}_L N_R x^0 \right],
\label{EqYukLep1}
\end{equation}
\begin{equation}
 f^{\prime}_{lR} \left[ \overline{e}_R N_{1L}(\cos(\theta_x) x_2^- + \sin(\theta_x) x_1^-) + \overline{\nu}_L N_{1R} x^0 \right],
\label{EqYukLep2}
\end{equation}
\begin{equation}
f_{s} s_1 (\overline{e}_L E_R^++ \overline{\nu}_L E_R^0),
\label{EqYukNu}
\end{equation}
and
\begin{equation}
f_{NE} (\cos(\theta_y) y_1^0 - \sin (\theta_y) y_2^0)\overline{N}_{1,2,3} E_R^0.
\end{equation}
There are potential collider constraints on the masses of the scalars, and of the vector-like fermions, however masses on order of a few hundred GeV for the scalars and on order of 500 GeV can be sufficient to avoid these constraints~\cite{acnscotowdm,Dermisek:2014qca}.  An interesting consequence of the proposed model is the existence of a particular mass scheme where the scalar DM species mass is on order of hundreds of MeV (the minimal Ma model can also accomodate MeV DM~\cite{Arhrib:2015dez}), but because of the multiple component nature the relic density constraints can still be met with the fermionic dark matter on order of 100 GeV.  However, in this mass range there are extra constraints that need to be considered.  In particular the Higgs invisible width contribution from the $ h \rightarrow s s$ is an important constraint, where the partial-width is given by
\begin{equation}
\Gamma_{h\rightarrow ss} = \frac{v^2 \lambda_{sh}^2 \sqrt{m_h^2 (m_h^2 -4 m_s^2}}{32 \pi m_h^4},
\end{equation}
which yields $\Gamma_{h\rightarrow ss}/\Gamma_h \approx 0.07$ for $m_s \approx 100$ MeV and $\lambda_{sh} = 0.01$.  Additionally, there are potential contraints from $ss \rightarrow L L$ via the $t$-channel exchange of $E_{R,L}$ as well as Big Bang Nucleosynthesis (BBN) constraints for such light DM~\cite{Henning:2012rm}.  In this mass range the scalar DM species only represents approximately one third of the total cosmological DM, for a particular set of masses and couplings, which reduces constraints on the annihilation cross section of DM.  Unlike other models with sub-GeV DM the $Z^{\prime}$ and the DM scalar DM species have couplings to SM particles and cannot avoid BBN constraints~\cite{Izaguirre:2015yja}, and so results from Planck on the annihilation cross-section rule out this particular mass scheme~\cite{planck2015}.

However, there are three additional mass schemes that produce thermal relic DM candidates in a mass range that can avoid the constraints from Planck:

\begin{itemize}
\item {\bf Model A:} $m_{E^0} = 455$ GeV, $m_{E^{\pm}} = 450$ GeV, $m_{F^0} = 600$ GeV,  $m_{x^{\pm}_1} =646$ GeV, $m_{x^{\pm}_2}=654$ GeV, $m_{x^0}=650$ GeV, $m_{y^{\pm}}= 247$ GeV, $m_{y^0_1}=250$ GeV, $m_{y^0_2}=252$ GeV, with $g_{\zeta} = 0.1$ and $m_{Z^{\prime}} = 1200 $ GeV.

\item {\bf Model B:} $m_{E^0} = 455$ GeV, $m_{E^{\pm}} = 450$ GeV, $m_{F^0} = 600$ GeV,  $m_{x^{\pm}_1} =646$ GeV, $m_{x^{\pm}_2}=654$ GeV, $m_{x^0}=650$ GeV, $m_{y^{\pm}}= 247$ GeV, $m_{y^0_1}=250$ GeV, $m_{y^0_2}=252$ GeV, with $g_{\zeta} = 0.00033$ and $m_{Z^{\prime}} = 2.7 $ GeV, $m_s = 100$ GeV, $m_{N_1}=70$ GeV.

\item {\bf Model C:} $m_{E^0} = 850$ GeV, $m_{E^{\pm}} = 825$ GeV, $m_{F^0} = 600$ GeV,  $m_{x^{\pm}_1} =998$ GeV, $m_{x^{\pm}_2}=1006$ GeV, $m_{x^0}=1002$ GeV, $m_{y^{\pm}}= 646$ GeV, $m_{y^0_1}=650$ GeV, $m_{y^0_2}=654$ GeV, with $g_{\zeta} = 0.0025$ and $m_{Z^{\prime}} = 20 $ GeV, $m_s=150$ GeV, $m_{N_1}=9$ GeV.

\end{itemize}
In each of these models the rest of the particle content is taken to be on the order of a TeV.  In order to determine relic denisty and direct detection constraints on the spin-independent (SI) cross section in an automated way, the relevant Lagrangian terms are implemented using FeynRules~\cite{feynrules} in order to generate model files to use with MicrOmegas~\cite{micromegas}.   For the numerical analysis the relic density is allowed to vary:
\begin{equation}
\label{relicExp}
0.11 \lesssim \Omega h^2 \lesssim 0.13.
\end{equation}
For direct detection the LUX 2016 and PandaX direct detection limits on the SI cross section are used (where the reported limit is on order of 4 times stronger than LUX 2015)~\cite{lux,lux2015,Tan:2016zwf}.  However, since this model is potentially multi-component, the direct detection constraints are fitted to the SI cross section modified by the proportion of the total relic density that each species contributes to the overall cosmological relic density for that particular mass~\cite{Bhattacharya:2013hva}.

\begin{figure}[htb]
\centering
\centering
\includegraphics[scale=1]{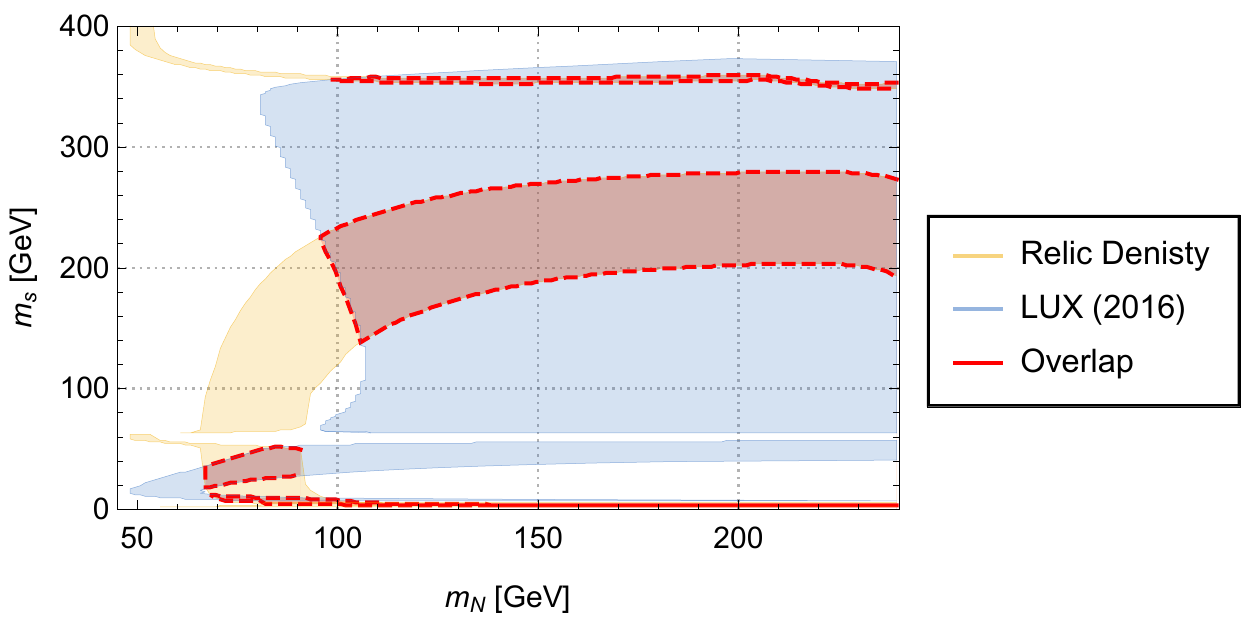}
\centering
\caption{Scan of Model A mass scheme in $m_s - m_N$ plane, where plotted points produce a DM relic density between $0.11 \lesssim \Omega_0 h^2 \lesssim 0.13$, are lower than the bound on $\sigma_{SI}$ from LUX 2016/PandaX, and the region where the DM candidates meet both constraints.}
\label{ModelA}
\end{figure}

\textit{Model A:}
 The dark matter masses are scanned over for the case where $f_s=0.22$, $f_{lL} = f_{lR} = 0.7$, and $\lambda_{sH} = 0.01$ and displayed in the $m_s - m_N$ mass plane in Fig.~\ref{ModelA}.  Even taking into account the latest LUX/PandaX constraints on the SI cross section, significant regions of the thermal relic DM parameter space survive.

\textit{Model B/C:} 
For these models a full numerical scan was not performed, however, both are able to fit the relic density range from Eq.~\ref{relicExp} and are below the SI cross section upper-bound from LUX/PanadaX, and for the mass choices are able to avoid the constraints on the annihilation cross section from Planck.  These parameter spaces are of potential interest for the existence of their relatively light $Z^{\prime}$ and, in Model C's case, their relatively heavy scalar masses.
\subsection{Collider Signatures}

\begin{figure}[htb]
\centering
\begin{minipage}{0.45\textwidth}
\centering
\includegraphics[scale=0.65,bb = 120bp 500bp 500bp 800bp,clip]{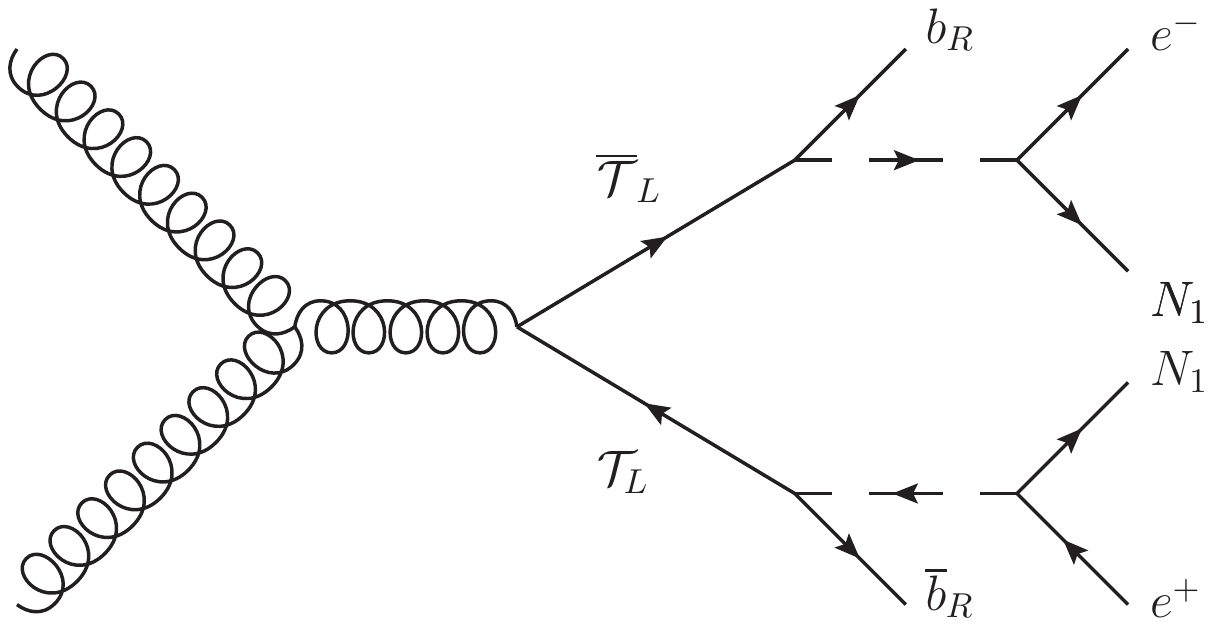}
\end{minipage}
\hfill

\begin{minipage}{0.45\textwidth}
\centering
\includegraphics[scale=0.65,bb = 120bp 500bp 500bp 800bp,clip]{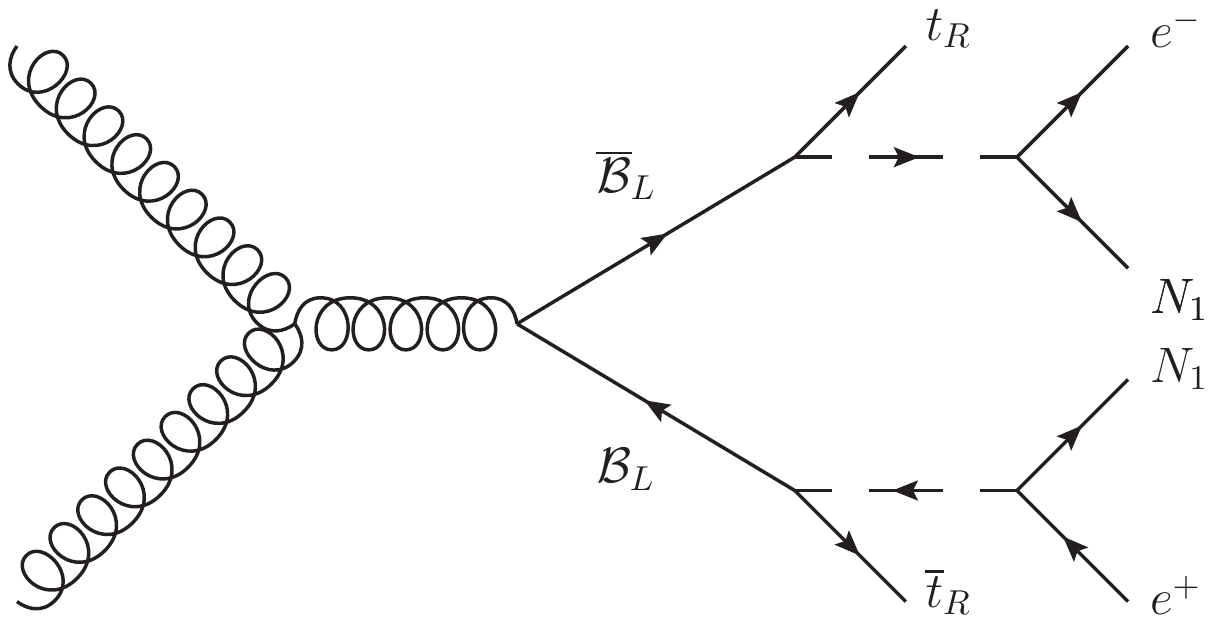}
\end{minipage}
\centering
\caption{Pair production from gluons of $\mathcal{T}$($\mathcal{B}$) and subsequent decay to SM particles and the fermionic DM species ($N_1$).}
\label{VLQcollider}
\end{figure}

The proposed model has additional EW states, an additional gauge boson, and additional heavy colored states any of which could produce a novel collider signature.  However, new EW scalar states that do not mix with the Higgs (ie inert or dark scalars) may be challenging to find at a hadron collider~\cite{acnscotowdm}, however the phenomneology of the $x_{1,2}^{\pm}$ states are essentially identical to previously studied models and so the primary source of potential collider signatures at the LHC are from the vector-like quarks or the $Z^{\prime}$.  The $\mathcal{U}$ and $\mathcal{D}$ vector-like quarks can be pair produced at the LHC, and their subsequent decay chain is to a quark, a lepton, and both species of DM, where the flavor of the lepton species will be fixed by the horizontal symmetry, however, this decay chain involves multiple mediators which can reduce the signal and complicate the analysis.  An alternative signature is the production and decay of $\mathcal{T}$ and $\mathcal{B}$, which decays to a top or bottom quark respectively, and $x^{\pm}$ which subsequently decays to a neutral fermion $N_i$ and a charged lepton as shown in Fig.~\ref{VLQcollider}.  Thus, if these vector-like quarks are the lightest colored-states, then the primary collider searches for colored particles in this model are to 2 bottom (top) quarks, 2 leptons, and missing energy,  in contradistinction to the typical vector-like quark searches~\cite{ATLAS:2016ovj}.  Addtionally, the $Z^{\prime}$ can be searched for at the LHC through dijet signatures~\cite{Chiang:2015ika,Fairbairn:2016iuf}, assuming the mass and coupling choices of Model A.  Future precision measurement of the Higgs coupling may also be able to rule out this model from either a more precise measurement of the invisible branching fraction~\cite{Ko:2016xwd} or from deviations in various Higgs boson couplings that occur in scotogenic models generally~\cite{seanma}.
\section{Conclusion}
In this work a scotogenic model of neutrino and charged lepton masses was extended to generate the first two generations of quark masses through their interaction with vector-like quarks.  Additionally, the binary tetrahedral symmetry $T^{\prime}$, in lieu of $A_4$, is utilized for the first time in the scotogenic framework.  Using a particular $T^{\prime}$ assignment which is softly-broken by the vector-like quark mass terms to a residual $Z_4$, the model is found to produce a mixing angle close to the Cabibbo angle.  The quark masses are generated through their interaction with the additional vector-like quarks, which carry both dark charge and are odd under an exactly conserved $Z_2$.  The first and second generation of quarks are allowed to transform under the $U(1)_D$, which necessitates the addition of $T^{\prime}$ singlets that are chiral under the dark gauge, but vector-like under the SM gauge group, in order to cancel anomalies. The model thus has a leptophobic $Z^{\prime}$, and two DM candidates (one scalar $s$ and one fermion $N$), and is found to successfully fit the relic density, the constraints on the annihilation cross section from Planck, and evade the latest SI cross section limits from the direct detection experiments LUX and PandaX.  In addition, the decays of the vector-like quarks $\mathcal{T}$ and $\mathcal{B}$ are shown to produce a very interesting signature which could be found at colliders in the future with an interesting final state when compared to existing vector-like quark models being investigated at the LHC.

{\bf Acknowledgments}
I am grateful to Anthony DiFranzo, Kiel Howe, Bithika Jain, Corey Kowanacki, Pyungwon Ko, Takaaki Nomura, Yusuke Shimizu and Peiwen Wu for critical discussions that made this work possible.
\bibliographystyle{utphys}
\bibliography{Bib}

\end{document}